\documentclass[a4paper, 12pt, oneside]{article}
\usepackage[cp1251]{inputenc}
\usepackage[english, russian]{babel}
\usepackage{ graphics,amsfonts, amsmath,bm,amssymb, array}
\usepackage[dvips]{graphicx}
\makeatletter
\usepackage{indentfirst}
\renewcommand{\baselinestretch}{1.2}
\usepackage[dvips]{graphicx}
\usepackage[flushleft,small]{caption2}

\renewcommand{\Re}{\mathop{\rm Re\,}}
\renewcommand{\Im}{\mathop{\rm Im\,}}
\renewcommand{\max}{\mathop{\rm max\,}}

\makeatother {\renewcommand{\baselinestretch}{1.09}

\begin{document}
\thispagestyle{empty} \large
\renewcommand{\abstractname}{\, }
\renewcommand{\refname}{\begin{center} REFERENCES\end{center}}
\thispagestyle{empty}
\large

\begin{flushright}\it\large
Dedicated to the Memory\\ of our Teachers\\ C. Cercignani and
K. Case
\end{flushright}

 \begin{center}
\bf Аналитическое решение второй задачи Стокса о поведении газа
над колеблющейся поверхностью. Часть II: математический аппарат для решения задачи
\end{center}\medskip
\begin{center}
  \bf V. A. Akimova\footnote{$vikont\_ava@mail.ru$},
  A. V. Latyshev\footnote{$avlatyshev@mail.ru$} and
  A. A. Yushkanov\footnote{$yushkanov@inbox.ru$}
\end{center}\medskip

\begin{center}
{\it Faculty of Physics and Mathematics,\\ Moscow State Regional
University, 105005,\\ Moscow, Radio str., 10--A}
\end{center}\medskip

\tableofcontents
\setcounter{secnumdepth}{4}

\begin{abstract}
In the present work the mathematical apparatus necessary for solving of second Stokes problem is developed. Second Stokes problem is the problem about
behaviour of rarefied gas filling half-space. A plane,
limiting half-space, makes harmonious oscillations in the eigen plane. 
At the heart of the analytical decision the homogeneous Riemann boundary value problem lays. The decision of homogeneous problem Римана is function, factorizing problem. Integral representations for factorizing function is deduced. With the help of factorizing function the factorization formula for dispersion function is proved. Zero of dispersion function are searched by means of the factorization formula of dispersion function.

В настоящей работе развивается математический аппарат, необходимый для
решения второй задачи Стокса. Вторая задача Стокса --- задача о
поведении разреженного газа, заполняющего полупространство. Плоскость,
ограничивающая полупространство, совершает гармонические колебания
в своей плоскости. Используется линеаризованное кинетическое уравнение. Рассматривается случай диффузного отражения молекул газа от стенки.
В основе аналитического решения лежит решение однородной задачи Римана из теории функций комплексного переменного. Подстановка общего решения задачи в виде разложения по собственным функциям характеристического уравнения в граничное условие на стенке приводит к сингулярному интегральному уравнению с ядром Коши. Затем сингулярное уравнение сводится к неоднородной краевой задаче. Решением однородной задачи Римана является функция, факторизующая коэффициент задачи. Для факторизующей функции и обратной к ней величине выводится ряд интегральных представлений. Эти интегральные представления используются в дальнейшем при нахождении скорости газа в полупространстве и вблизи границы, при построении функции распределения газовых молекул, при нахождении силы трения газа на границу и диссипации энергии колеблющейся границы. С помощью факторизующей функции доказывается формула факторизации дисперсионной функции. С помощью формулы факторизации в явном виде находятся нули дсперсионной функции.


\end{abstract}

\begin{center}
\item{}\section{Введение}
\end{center}

Задача о поведении газа над движущейся поверхностью в последние годы
привлекает пристальное внимание \cite{Stokes} -- \cite{15}. Это связано
с развитием современных технологий, в частности, технологий наноразмеров.
В \cite{Yakhot} -- \cite{15} эта задача решалась численными или
приближенными методами. В настоящей работе показано, что эта задача
допускает аналитическое решение. Аналитическое решение строится с помощью теории обобщенных функций и сингулярных интегральных уравнений.

Впервые задача о поведении газа над стенкой, колеблющейся в своей плоскости,
была рассмотрена Дж. Г. Стоксом \cite{Stokes}. Задача решалась гидродинамическим методом без учёта эффекта скольжения. Обычно такую задачу называют второй задачей Стокса \cite{Yakhot}--\cite{SS-2002}.

В последние годы на тему этой задачи появился ряд публикаций. В работе
\cite{Yakhot} рассматривается бесконечная колеблющаяся поверхность. Задача рассматривается для любых частот колебания поверхности. Из кинетического уравнения БГК получено уравнение типа гидродинамического. Рассматриваются гидродинамические граничные условия. Вводится коэффициент, связывающий скорость газа на поверхности со скоростью поверхности. Показано, что в случае высокочастотных колебаний сила трения, действующая на поверхность, не зависит от частоты.
В работе \cite{SK-2007} получены коэффициенты вязкостного и
теплового скольжения с использованием различных модельных уравнений. Использованы как максвелловские граничные условия, так и граничные условия Черчиньяни --- Лэмпис.

Второй задаче Стокса посвящена диссертация \cite{15}.
Наиболее близкая к решенной в первой и второй главах диссертации \cite{15} задача решена в статье \cite{10}: рассматривается газовый поток над бесконечной пластиной, совершающей гармонические колебания в собственной плоскости. Найдена скорость газа над поверхностью и сила, действующая на поверхность со стороны газа. Для случая низких частот задача решена на основе уравнения Навье --- Стокса. Для произвольных скоростей колебаний поверхности задача решена численными методами на основе кинетического уравнения Больцмана с интегралом столкновений в форме БГК (Бхатнагар, Гросс, Крук). 

Работа \cite{11} является экспериментальным исследованием. Эксперименты показывают, что при низких частотах колебаний резонатора, действующая на него со стороны газа сила трения прямо пропорциональна частоте колебания резонатора. При высоких частотах колебания резонатора ($~10^8$ Гц) действующая на него сила трения от частоты колебаний не зависит.

В последнее время задача о колебаниях плоской поверхности в собственной
плоскости изучается и для случая неньютоновских жидкостей \cite{5} и
\cite{6}.

В статье \cite{12} рассматривается пример практического применения
колебательной системы, подобной рассматриваемой во второй задаче Стокса, в области нанотехнологий.

В настоящей работе вторая задача Стокса впервые решается аналитически.
При этом используются сингулярные интегральные уравнения с ядром Коши и обобщенные функции.

В п. 3 решается однородная краевая задача Римана, лежащая в основе
аналитического решения кинетического уравнения. Ищется функция $X(z)$, аналитическая в комплексной плоскости с разрезом вдоль действительной положительной полуоси, такая, что ее граничные значения сверху и снизу на действительной положительной полуоси связаны условием линейного сопряжения:
$$
X^+(\mu)=G(\mu)X^-(\mu), \qquad \mu>0.
$$

Решение этой задачи --- функция $X(z)$ называется факторизующей функцией
(коэффициент задачи). Именно эта функция позволяет факторизовать
дисперсионную функцию. Решение задачи Римана находится в обоих случаях: когда индекс задачи равен нулю и индекс задачи равен единице. В первом случае $X(z)$ ищется в классе ограниченных в бесконечно удаленной точке функций в первом случае и в классе исчезающих в бесконечно удаленной точке функций во втором случае.

В п. 4 выводятся необходимые в дальнейшем интегральные представления как для факторизующей функции $X(z)$, так и для функции $1/X(z)$. Эти интегральные представления восстанавливают значение функции в комплексной плоскости с разрезом вдоль положительной действительной полуоси через их граничные значения, взятые на разрезе.

В п. 5 доказываются формулы факторизации дисперсионной функции в виде произведения $X(z)X(-z)$ с точностью до некоторого множителя.
Факторизация дисперсионной функции позволяет в явном виде найти нули дисперсионной функции.

Математический аппарат развивается для решения граничных задач для уравнения:
$$
\mu\dfrac{\partial h}{\partial x_1}+z_0h(x_1,\mu)=\dfrac{1}{\sqrt{\pi}}
\int\limits_{-\infty}^{\infty}\exp(-{\mu'}^2)h(x_1,\mu')d\mu',\ 
\eqno{(1.1)}
$$
где
$$
z_0=1-i\omega_1.
$$
Именно к уравнению (1.1) и сводится основное уравнение из второй задачи
Стокса \cite{ALY-2011}.
\begin{center}
\item{}\section{Предшествующие результаты}
\end{center}

Разделение переменных в уравнении (1.1) осуществляется следующей подстановкой
$$
h_\eta(x_1,\mu)=\exp\Big(-\dfrac{x_1z_0}{\eta}\Big)\Phi(\eta,\mu),
\eqno{(2.1)}
$$
где $\eta$ -- параметр разделения, или спектральный параметр,
вообще говоря, комплексный.

Подставляя (2.1) в уравнение (1.1) получаем характеристическое уравнение
$$
(\eta-\mu)\Phi(\eta,\mu)=\dfrac{\eta}{\sqrt{\pi}z_0}
\int\limits_{-\infty}^{\infty}
\exp(-{\mu'}^2)\Phi(\eta,\mu')d\mu'.
\eqno{(2.2)}
$$
Если ввести обозначение
$$
n(\eta)=\dfrac{1}{z_0}\int\limits_{-\infty}^{\infty}
\exp(-{\mu'}^2)\Phi(\eta,\mu')d\mu',
\eqno{(2.3)}
$$
то уравнение (2.2) может быть записано с помощью (2.3) в виде
$$
(\eta-\mu)\Phi(\eta,\mu)=\dfrac{1}{\sqrt{\pi}}\eta n(\eta),\qquad
\eta\in \mathbb{C}.
\eqno{(2.4)}
$$
Уравнение (2.4) является конечным (недифференциальным) уравнением.
Условие (2.3) называется нормировочным условием,
нормировочным интегралом, или просто нормировкой.

Решение характеристического уравнения для действительных значений
параметра $\eta$ будем искать в пространстве обобщенных функций \cite{6}.
Обобщенное решение уравнения (2.4) имеет вид:
$$
\Phi(\eta,\mu)=\dfrac{1}{\sqrt{\pi}}\eta P\dfrac{1}{\eta-\mu}+
\exp(\eta^2)\lambda(\eta)\delta(\eta-\mu),
\eqno{(2.5)}
$$
где $-\infty<\eta, \mu <+\infty$, $\delta(x)$ -- дельта--функция Дирака, символ $Px^{-1}$
означает главное значение интеграла при интегрировании $x^{-1}$,
$\lambda(z)$ -- дисперсионная функция, введенная равенством
$$
\lambda(z)=1-i\omega_1+\dfrac{z}{\sqrt{\pi}}\int\limits_{-\infty}^{\infty}
\dfrac{\exp(-\tau^2)d\tau}{\tau-z}.
$$
Эту функцию можно преобразовать к виду: $\lambda(z)=-i\omega_1+\lambda_0(z)$,
где $\lambda_0(z)$ -- известная функция из теории плазмы,
$$
\lambda_0(z)=\dfrac{1}{\sqrt{\pi}}\int\limits_{-\infty}^{\infty}
\dfrac{e^{-\tau^2}\tau d\tau}{\tau-z}.
$$
Собственные функции (2.5) определены с использованием нормировки
$$
n(\eta)\equiv 1.
$$
В силу однородности уравнения (1.1) можно считать, что
$
n(\eta)\equiv 1.
$

Собственные функции (2.5) называются собственными функциями непрерывного
спектра, ибо спектральный параметр $\eta$ непрерывным образом заполняет
всю действительную прямую.

Таким образом, собственные решения уравнения (1.1) имеют вид
$$
h_\eta(x,\mu)=\exp\Big(-\dfrac{x_1}{\eta}z_0\Big)
\Big[\dfrac{1}{\sqrt{\pi}}\eta P\dfrac{1}{\eta-\mu}+
\exp(\eta^2)\lambda(\eta)\delta(\eta-\mu)\Big].
\eqno{(2.6)}
$$

Собственные решения (2.6) отвечают непрерывному спектру характеристического уравнения, ибо спектральный параметр непрерывным образом пробегает всю числовую прямую, т.е. непрерывный спектр $\sigma_c$
есть вся конечная часть числовой прямой: $\sigma_c=(-\infty,+\infty)$.

По условию задачи мы ищем решение, невозрастающее вдали от стенки.
Поэтому далее будем рассматривать положительную часть непрерывного спектра.
В этом случае собственные решения (2.6) являются исчезающими вдали от стенки. В связи с этим спектром граничной задачи будем называть положительную действительную полуось параметра $\eta$:
$\sigma_c^{\rm problem}=(0,+\infty)$.

Приведем необходимые в дальнейшем формулы Сохоцкого для дисперсионной функции:
$$
\lambda^{\pm}(\mu)=\pm i s(\mu)-i\omega_1+\lambda_0(\mu),\qquad
s(\mu)=\sqrt{\pi}\mu e^{-\mu^2}.
$$
Разность граничных значений дисперсионной функции отсюда равна:
$$
\lambda^+(\mu)-\lambda^-(\mu)=2s(\mu)i,
$$
полусумма граничных значений равна:
$$
\dfrac{\lambda^+(\mu)+\lambda^-(\mu)}{2}=\lambda(\mu)=
-i\omega_1+\lambda_0(\mu).
$$

Заметим, что на действительной оси действительная часть
дисперсионной функции $\lambda_0(\mu)$ имеет два нуля $\pm\mu_0$, $\mu_0=0.924\cdots$. Эти два нуля в силу четности функции $\lambda_0(\mu)$ различаются лишь знаками.

Отметим, что на действительной оси дисперсионную функцию удобнее использовать в численных расчетах в виде (см. \cite{19})
$$
\lambda_0(\mu)=1-2\mu^2 \int\limits_{0}^{1}\exp(-\mu^2(1-t^2))dt,\qquad
\mu\in(-\infty,+\infty).
$$

Разложим дисперсионную функцию в ряд Лорана по отрицательным степеням
переменного $z$ в окрестности бесконечно удаленной точки:
$$
\lambda(z)=-i\omega_1-\dfrac{1}{2z^2}-\dfrac{3}{4z^4}-\dfrac{15}{8z^6}-\cdots,
\quad z\to \infty.
\eqno{(2.7)}
$$

Из разложения (2.7) видно, что при малых значениях $\omega_1$
дисперсионная функция имеет два отличающиеся лишь знаками комплексно--значных нуля:
$$
\pm\eta_0^{(0)}(\omega_1)=\dfrac{1+i}{2\sqrt{\omega_1}}.
$$

Отсюда видно, что при $\omega_1\to 0$ оба нуля дисперсионной функции
имеют пределом одну бесконечно удаленную точку $\eta_i=\infty$ кратности (порядка) два.

Из разложения (2.7) видно так же, что значение дисперсионной функции в
бесконечно удаленной точки равно:
$$
\lambda(\infty)=-i\omega_1.
$$

В \cite{ALY-2011} показано, что число нулей дисперсионной функции равно удвоенному индексу коэффициента задачи
$$
N=2\varkappa(G),
$$
где $\varkappa=\varkappa(G)$ -- индекс функции $G(t)$ -- число
оборотов кривой $\Gamma(\omega_1)$ относительно начала координат, совершаемых в положительном направлении, $G(\mu)=\lambda^+(\mu)/\lambda^-(\mu)$

Введем угол $ \theta( \mu)= \arg G(\mu)$ --
главное значение аргумента функции $G(\mu)$, фиксированное
в нуле условием $ \theta(0)=0$.

Введем выделенную частоту колебаний пластины, ограничивающей газ:
$$
\omega_1^*=\max\limits_{0<\mu<+\infty}
\sqrt{-\lambda_0^2(\mu)+s^2(\mu)}\approx 0.733.
$$

Эту частоту колебаний называется {ALY-2011} {\it критической}.

Показано \cite{ALY-2011}, что в случае, когда частота колебаний пластины меньше критической, т.е. при $0\leqslant \omega <\omega_1^*$, индекс функции $G(t)$ равен единице. Это означает, что число комплексно--значных нулей дисперсионной функции в разрезанной комплексной плоскости с разрезом вдоль действительной оси, равно двум.

В случае, когда частота колебаний пластины превышает критическую
($\omega>\omega_1^*$) индекс функции $G(t)$ равен нулю: $\varkappa(G)=0$. Это означает, что дисперсионная функция не имеет нулей в верхней и нижней полуплоскостях. В этом случае дискретных (частных) решений исходное кинетическое уравнение (1.1) не имеет.

При $0 \leqslant \omega_1<\omega_1^*$ нули дисперсионной функции
обозначим через $\eta_0(\omega_1)$ и $-\eta_0(\omega_1)$. В силу четности дисперсионной функции ее конечные нули различаются только знаками, имея одинаковые модули.

Таким образом, дискретный спектр характеристического уравнения,
состоящий из нулей дисперсионной функции, в случае $0\leqslant \omega_1<\omega_1^*$ есть множество из двух точек $\sigma_d(\omega_1)=\{\eta_0(\omega_1),
-\eta_0(\omega_1)\}$. При $\omega_1>\omega_1^*$ дискретный спектр ---
это пустое множество. При $0\leqslant \omega_1<\omega_1^*$ собственными функциями характеристического уравнения являются следующие два решения характеристического уравнения:
$$
\Phi(\pm \eta_0(\omega_1),\mu)=
\dfrac{1}{\sqrt{\pi}}\dfrac{\pm \eta_0(\omega_1)}{\pm \eta_0(\omega_1)-\mu}
$$
и два соответствующих собственных решения характеристического уравнения:
$$
h_{\pm \eta_0(\omega_1)}(x_1,\mu)=
\exp \Big(-\dfrac{x_1z_0}{\pm \eta_0(\omega_1)}\Big)\dfrac{1}{\sqrt{\pi}}\dfrac{\pm \eta_0(\omega_1)}{\pm \eta_0(\omega_1)-\mu}.
$$

Под $\eta_0(\omega_1)$ будем понимать тот из нулей
дисперсионной функции, который обладает свойством:
$$
\Re \dfrac{1-i\omega_1}{\eta_0(\omega_1)}>0.
$$
Для этого нуля убывающее собственное решение кинетического
уравнения (1.1) имеет вид
$$
h_{\eta_0(\omega_1)}(x_1,\mu)=\dfrac{1}{\sqrt{\pi}}
\exp\Big(-\dfrac{x_1z_0}{\eta_0(\omega_1)}\Big)\dfrac{\eta_0(\omega_1)}
{\eta_0(\omega_1)-\mu}.
$$
Это означает, что дискретный спектр рассматриваемой граничной
задачи состоит из одной точки $\sigma_d^{\rm problem}=\{\eta_0(\omega_1)\}$ в случае $0 <\omega_1<\omega_1^*$. При $\omega_1\to 0$ оба нуля, как уже указывалось выше, перемещаются в одну и ту же бесконечно удаленную точку. Это значит, что в этом случае дискретный спектр характеристического уравнения состоит из одной бесконечно удаленной точки кратности два:
$\sigma_d(0)=\eta_i=\infty$ и является присоединенным к
непрерывному спектру. Этот спектр является также и спектром рассматриваемой граничной задачи. Однако, в этом случае дискретных (частных) решения ровно два:
$$
h_1(x_1,\mu)=1, \qquad h_2(x_1,\mu)=x_1-\mu.
$$

\begin{center}
\item\section{Однородная краевая задача Римана}
\end{center}

В основе аналитического решения граничных задач
кинетической теории лежит решение однородной краевой задачи
Римана (см. \cite{Gakhov}) с
коэффициентом $G(\mu)=\lambda^+(\mu)/\lambda^-(\mu)$:
$$
\dfrac{X^+( \mu)}{X^-( \mu)}=G(\mu),\;\qquad \mu>0,
$$
или
$$
\dfrac{X^+( \mu)}{X^-( \mu)}=\dfrac{ \lambda^+( \mu)}{ \lambda^-( \mu)},
\qquad \mu>0.
\eqno{(3.1)}
$$

Однородная краевая задача Римана (3.1) называется также (см. \cite{Gakhov}) задачей факторизации коэффициента $G(\mu)$.

Задача (3.1) означает, что отношение $\lambda^+( \mu)/ \lambda^-( \mu)$
можно заменить отношением функций $X^+( \mu)/X^-( \mu)$,
являющихся граничными значениями функции $X(z)$, аналитической в
комплексной плоскости $\mathbb{C}$ и имеющей в качестве линии
скачков положительную действительную полуось. Дисперсионная
функция имеет в качестве линии скачков всю действительную ось.

Итак, в задаче Римана (3.1) ищется функция $X(z)$, аналитическая в
комплексной плоскости с разрезом вдоль действительной положительной полуоси. При этом в первом регионе изменения параметра $\omega_1$
($\omega_1\in [0,\omega_1^*$)) функция $X(z)$ ищется в классе
аналитических в области $\mathbb{C}\setminus \mathbb{R}_+$, исчезающих в бесконечно удаленной точке. Во втором регионе (когда $\omega_1\in (\omega_1^*,+\infty)$)) функция X(z) ищется в классе аналитических в области $\mathbb{C}\setminus \mathbb{R}_+$, ограниченных в бесконечно удаленной точке.

Сначала рассмотрим случай, когда индекс задачи равен единице.

Введем регулярную ветвь аргумента $\theta(\mu)=\arg G(\mu)$,
фиксированную в нуле условием $\theta(0)=0$. Перепишем краевое условие (3.2) в виде
$$
\dfrac{X^+( \mu)}{X^-( \mu)}=
\big|G(\mu)\big|\exp\Big(i[\theta(\mu)+2\pi k]\Big),\; k=0,\pm 1, \pm 2,\dots , \;\quad \mu>0.
$$

Логарифмируя это краевое условие, получаем счетное семейство краевых задач $$
\ln X^+( \mu)-\ln X^-( \mu)=$$$$=\ln|G(\mu)|+ i( \theta( \mu)+2 \pi k), \qquad
k=0,\pm 1,\pm 2,\cdots , \; \mu>0.
\eqno{(3.3)}
$$
Учитывая, что приращение аргумента $\theta(\mu)$ на полуоси $[0,+\infty]$
в рассматриваемом регионе изменения параметра задачи $\omega_1$ равно $2\pi$, возьмем в краевом условии (3.3) $k=-1$. Теперь получаем конкретное краевое условие:
$$
\ln X^+( \mu)-\ln X^-( \mu)=\ln|G(\mu)|+ i( \theta( \mu)-2\pi),\qquad
\mu>0.
\eqno{(3.4)}
$$

Решение этой задачи, как задачи определения аналитической функции по скачку, имеет вид
$$
\ln X(z)= \dfrac{1}{2\pi i} \int\limits_{0}^{ \infty}
\dfrac{\ln|G(\mu)|+i[ \theta( \mu)-2\pi]}{ \mu-z}\,d \mu.
\eqno{(3.5)}
$$

Обозначим далее
$$
V(z)= \dfrac{1}{2\pi i} \int\limits_{0}^{\infty}
\dfrac{\ln|G(u)|+i\zeta(u)}{u-z}du.
\eqno{(3.6)}
$$
где
$$
\zeta(u)=\theta(\tau)-2\pi.
$$
Тогда из равенства (3.5) и (3.6) имеем:
$$
X(z)= \exp V(z).
\eqno{(3.7)}
$$

Выясним поведение функции $X(z)$ в начале координат. При $z \to 0$ имеем:
$$
V(z)=- \dfrac{ \theta(0)- 2\pi}{2 \pi}\ln z +O_1(z) \quad
(z \to 0),
$$
где $ \theta(0)=0$, а $O_1(z)$ -- ограниченная функция в начале координат.

Теперь видно, что функция $X(z)=z \exp O_1(z)$
является исчезающей в начале 
координат.

Чтобы сделать решение задачи (3.7) неисчезающим в нуле,
переопределим решение (3.7) следующим образом:
$$
X(z)= \dfrac{1}{z} \exp V(z),
\eqno{(3.8)}
$$
где функция $V(z)$ определяется равенством (3.6).

\begin{figure}[h]
\begin{center}
\includegraphics[width=15.0cm, height=12cm]{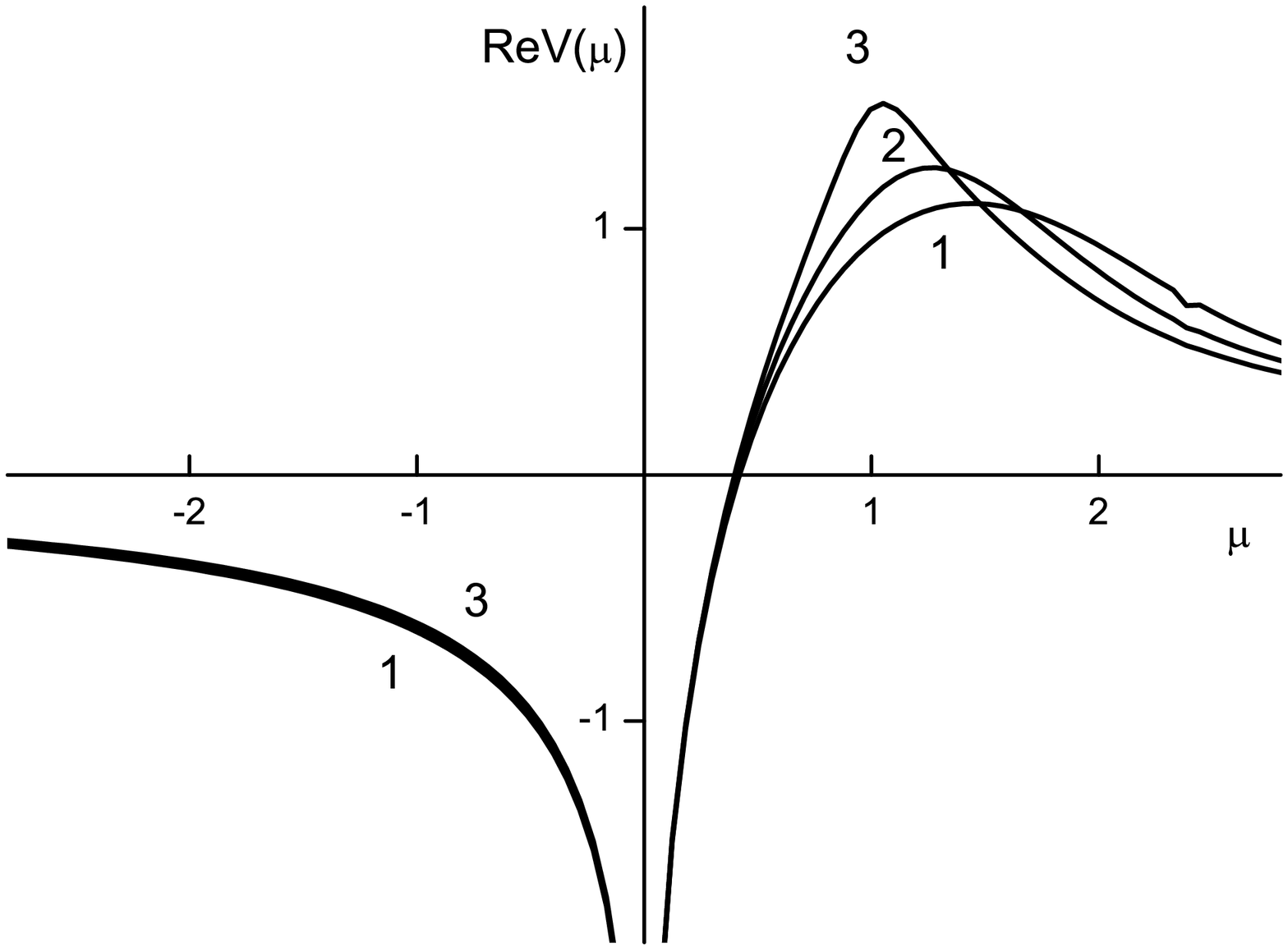}
\end{center}
\begin{center}
{{Рис. 1. Поведение функции $\Re V(\mu)$ на действительной оси. Кривые 
$1,2,3$ отвечают соответственно значениям $\omega_1=0.1,0.3,0.5$.}}
\end{center}
\end{figure}
\begin{figure}[h]
\begin{center}
\includegraphics[width=15.0cm, height=12cm]{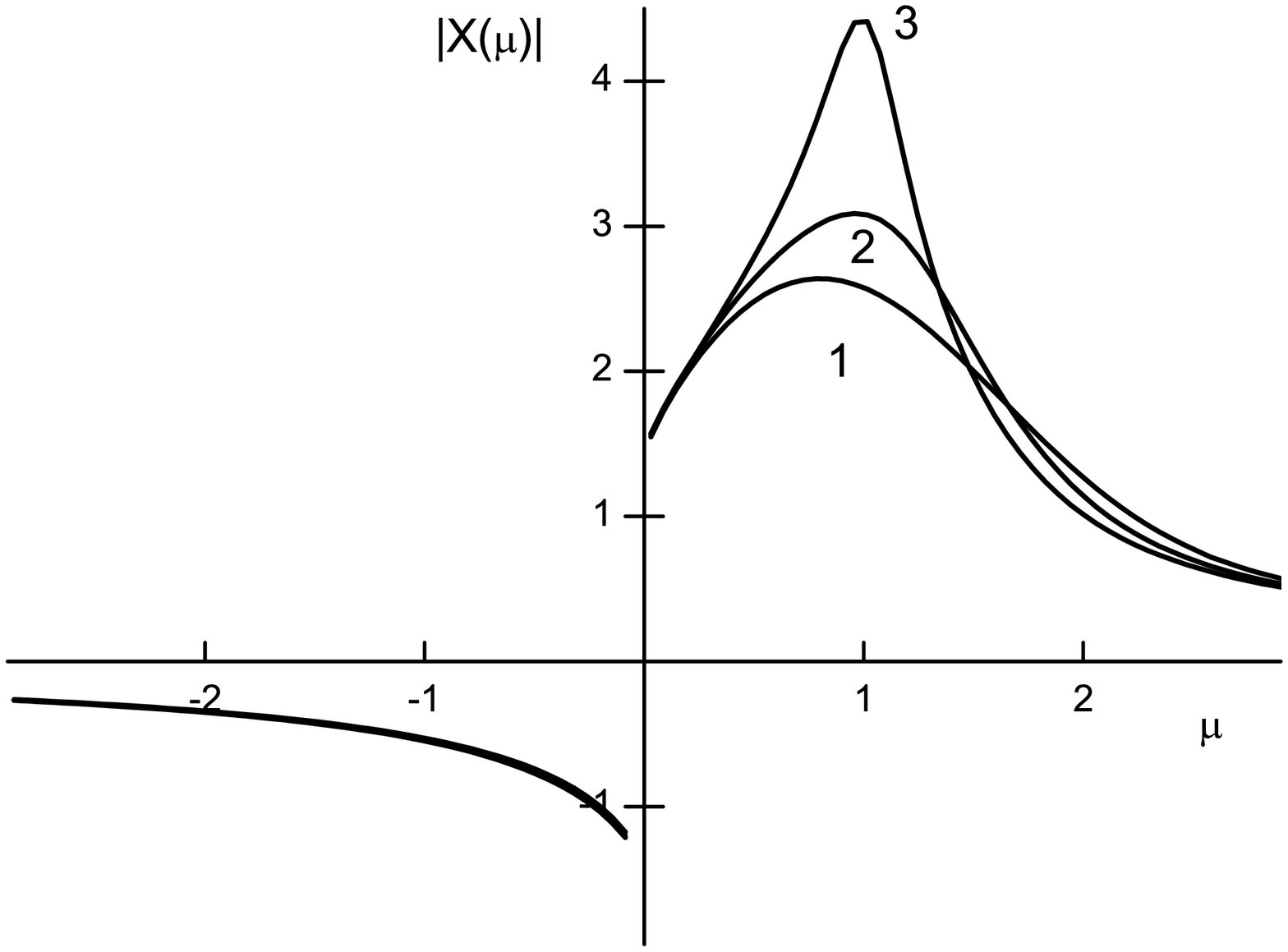}
\end{center}
\begin{center}
{{Рис. 2. Поведение функции $|X(\mu)|=\dfrac{1}{\mu}\exp(\Re V(\mu))$ на действительной оси. Кривые
$1,2,3$ отвечают соответственно значениям $\omega_1=0.1,0.3,0.5$.}}
\end{center}
\end{figure}

Такое переопределение функции $X(z)$ возможно, ибо решением
задачи (3.1) является любая функция вида $f(z)\exp V(z)$,
где $f(z)$ -- произвольная функция, аналитическая везде в
комплексной плоскости, кроме, быть может, точки $z=0$ ---
левого конца промежутка интегрирования, входящего в конструкцию функции $V(z)$.

Итак, решение (3.8) задачи (3.1) есть функция, ограниченная в окрестности
начала координат и исчезающая при $z\to \infty$.

Эту функцию далее будем называть факторизующей функцией,
ибо она осуществляет факторизацию коэффициента задачи $ G( \mu)$.

Согласно формулам Сохоцкого для функции $V(z)$ сверху и снизу на
положительной действительной полуоси имеем:
$$
V^{\pm}(\mu)=\pm \dfrac{1}{2}\Big[\ln|G(\mu)|+i \zeta(\mu)\Big]+V(\mu),
\qquad \mu>0,
\eqno{(3.9)}
$$
где
$$
V(\mu)=\dfrac{1}{2\pi i}\int\limits_{0}^{\infty}
\dfrac{\ln|G(\tau)|+i\zeta(\tau)}{\tau-\mu}\,d\tau,
$$
$V(\mu)$ -- сингулярный интеграл при $\mu>0$, имеющий в
точке $\mu=0$ бесконечный разрыв.

На основании (3.9) при $\mu>0$ имеем
$$
X^+(\mu)-X^-(\mu)= \dfrac{1}{\mu}\exp{V^+(\mu)}-\dfrac{1}{\mu}\exp{V^-(\mu)}=
$$
$$
=\dfrac{1}{\mu}\exp{V(\mu)}
\Big\{e^{\Theta(\mu)}-e^{-\Theta(\mu)}\Big\}=
2X(\mu)\sh\Theta(\mu),
\eqno{(3.10)}
$$
где
$$
\Theta(\mu)=\dfrac{1}{2}\Big[\ln|G(\mu)|+i \zeta(\mu)\Big].
$$

\begin{figure}[h]
\begin{center}
\includegraphics[width=15.0cm, height=12cm]{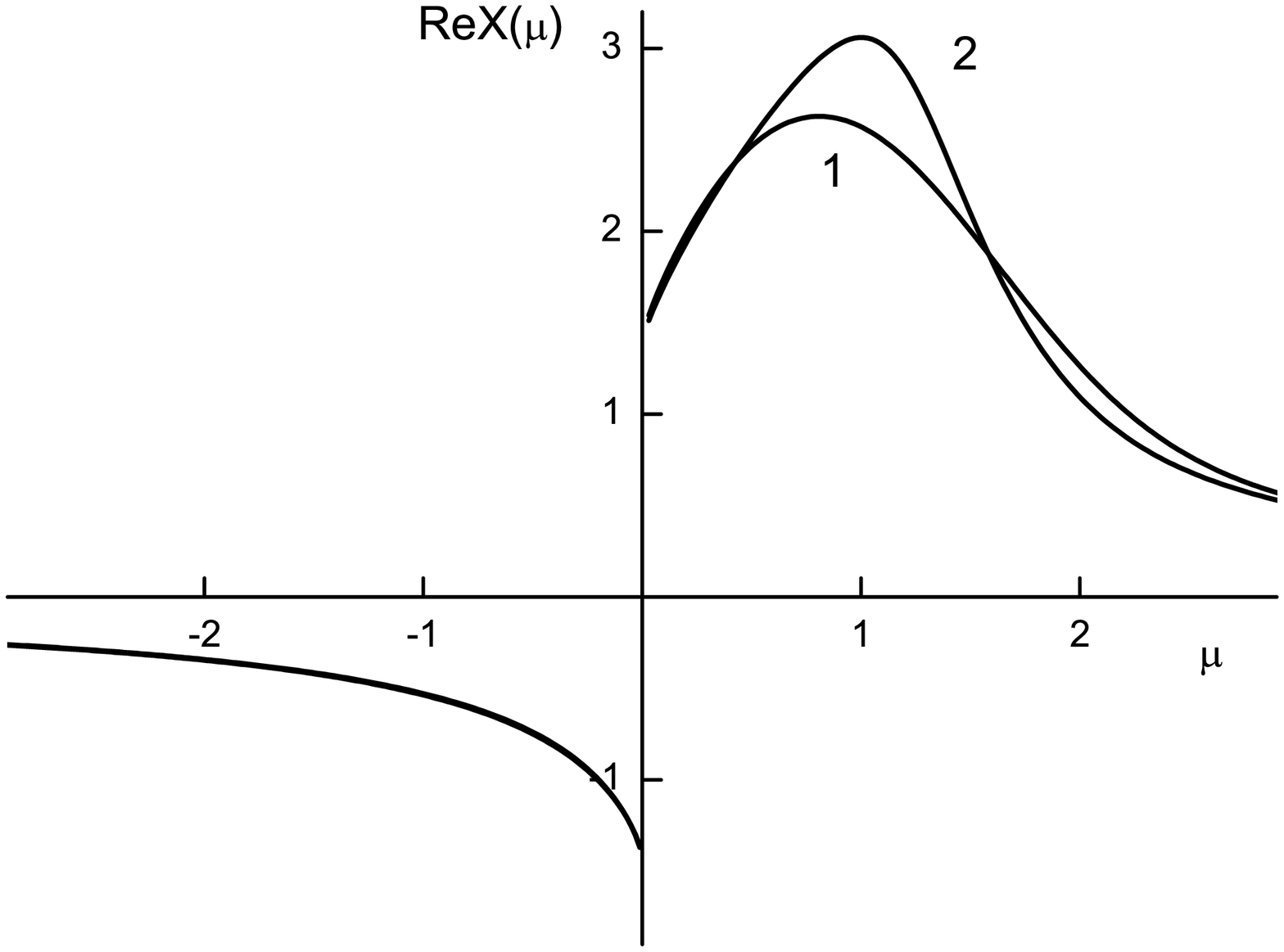}
\end{center}
\begin{center}
{{Рис. 3. Поведение функции 
$\Re X(\mu)=\dfrac{1}{\mu}\exp(\Re V(\mu))\cos(\Im V(\mu))$ на действительной оси. Кривые
$1,2$ отвечают соответственно значениям $\omega_1=0.1,0.3$.}}
\end{center}
\end{figure}

Аналогично получаем, что
$$
\dfrac{1}{X^+(\mu)}-\dfrac{1}{X^-(\mu)}=
\mu \exp(-V(\mu))\big[\exp(-\Theta(\mu))-\exp(\Theta(\mu))\big]=
$$
$$
=-2\dfrac{\sh\Theta(\mu)}{X(\mu)}.
\eqno{(3.11)}
$$

Аналогично (3.10) и (3.11) получаем
$$
\dfrac{X^+(\mu)+X^-(\mu)}{2}=X(\mu)\ch\Theta(\mu), \qquad
\mu>0,
\eqno{(3.12)}
$$
$$
\dfrac{1}{2}
\left[\dfrac{1}{X^+(\mu)}+\dfrac{1}{X^-(\mu)}\right]=
\dfrac{\ch\Theta(\mu)}{X(\mu)}, \qquad \mu>0.
\eqno{(3.13)}
$$

Подчеркнем, что формулы (3.10)--(3.13) справедливы на
положительной действительной полуоси $\mu>0$.

Теперь рассмотрим случай, когда индекс задачи равен нулю. В этом случае
$\omega_1\in (\omega_1^*,+\infty)$.
Приращение угла $\theta(\mu)$ на положительной полуоси равно нулю и решение однородной краевой задачи Римана дается формулой (3.7)
$$
X(z)= \exp V(z),
\eqno{(3.7)}
$$
в которой
$$
V(z)=\dfrac{1}{2\pi i}\int\limits_{0}^{\infty}\dfrac{\ln G(\tau)d\tau}
{\tau-z},
$$
а под $\ln G(\tau)$ понимается такая ветвь логарифма,
которая фиксирована в нуле условием: $\ln G(0)=0$. Отметим, что в силу свойств угла $\theta(\mu)$, имеем: $\ln G(\infty)=0$.

Выпишем формулы Сохоцкого для граничных значений сверху и снизу
на положительной действительной полуоси функции $V(z)$:
$$
V^{\pm}(\mu)=\pm \Theta(\mu)+V(\mu),
$$
где
$$
\Theta(\mu)=\dfrac{1}{2}\ln G(\mu), \qquad
V(\mu)=\dfrac{1}{2\pi i}\int\limits_{0}^{\infty}\dfrac{\ln G(\tau)d\tau}
{\tau-\mu}.
$$

В заключение приведем формулы для вычисления угла
$\theta(\mu)=\arg G(\mu)$. В первом регионе $\omega_1\in [0,\omega_1^*)$,
когда индекс задачи равен единице, угол вычисляется по формуле:
$$
\theta(\mu)=\left\{\begin{array}{c}
\arcctg\dfrac{\lambda_0^2(\mu)+\omega_1^2-s^2(\mu)}{2\lambda_0(\mu)s(\mu)},
\qquad \lambda_0(\mu)>0, \\
\arcctg\dfrac{\lambda_0^2(\mu)+\omega_1^2-s^2(\mu)}{2\lambda_0(\mu)s(\mu)}+\pi,
\quad\lambda_0(\mu)\leqslant 0.
\end{array}\right.
$$

Во втором регионе $\omega_1\in (\omega_1^*,+\infty)$, индекс задачи равен нулю, а угол вычисляется по формуле:
$$
\theta(\mu)=
\arctg\dfrac{2\lambda_0(\mu)s(\mu)}{\lambda_0^2(\mu)+\omega_1^2-s^2(\mu)}.
$$

\begin{center}
\item \section{Интегральное представление факторизующей функции}
\end{center}

Выведем интегральное представление для функции
$X(z)$, факторизующей коэффициент краевой задачи Римана.

Начнем со случая $\varkappa(G)=1$.

Функция $X(z)$ аналитична везде в замкнутой комплексной плоскости, за исключением разреза вдоль $\Delta^+=[0,+\infty]=\mathbb{\bar R_+}$.
Возьмем замкнутый контур $\Gamma_\varepsilon$,
изображенный на рис. 4.1, радиус $R$ большой окружности этого контура равен $R=1/ \varepsilon , \varepsilon>0, \varepsilon$
 -- достаточно малое положительное число, радиус $r$
 малой окружности равен $r= \varepsilon$.

Возьмем произвольную точку
$z \in \mathbb{ \overline{C}} \setminus \Delta^+$. Число
$R=1/ \varepsilon$ выберем столь большим,
чтобы эта точка лежала внутри замкнутого контура
$ \Gamma_ \varepsilon$. Согласно интегральной формуле Коши
для функции $X(z)$ справедливо интегральное представление
$$
X(z)= \dfrac{1}{2 \pi i}\oint\limits_{ \Gamma_ \varepsilon}
\dfrac{X(z')}{z'-z}\,d z'.
\eqno{(4.1)}
$$
В силу аддитивности интеграла имеем:
$$
X(z)= \dfrac{1}{2 \pi i} \left[ \int\limits_{AB}^{}+
\int\limits_{BCD}^{}+\int\limits_{DE}^{}+
\int\limits_{EFA}^{}\right] \dfrac{X(z')}{z'-z}\, dz'.
\eqno{(4.2)}
$$
Левая часть этого равенства не зависит от
$ \varepsilon$, следовательно, от $\varepsilon$ не зависит и
правая часть. Вычислим предел правой части равенства
(4.2) при $ \varepsilon \to 0$. Прежде всего имеем:
$$
\lim_{\varepsilon \to 0} \int\limits_{AB}^{}
\dfrac{X(z')}{z'-z}\,dz'= \int\limits_{0}^{+\infty}
\dfrac{X^+(u)}{u-z}\,du,
\eqno{(4.3)}
$$
$$
\lim_{ \varepsilon \to 0} \int\limits_{DE}^{}
\dfrac{X(z')}{z'-z}\,d z'=-\int\limits_{0}^{+ \infty}
\dfrac{X^-(u)}{u-z}\,du.
\eqno{(4.4)}
$$
Покажем теперь, что
$$
\lim_{ \varepsilon \to 0} \int\limits_{EFA}^{}
\dfrac{X(u')}{u'-z}\,du'=0,
\eqno{(4.5)}
$$

\begin{figure}[th]
\begin{center}
\includegraphics[width=18.0cm, height=24cm]{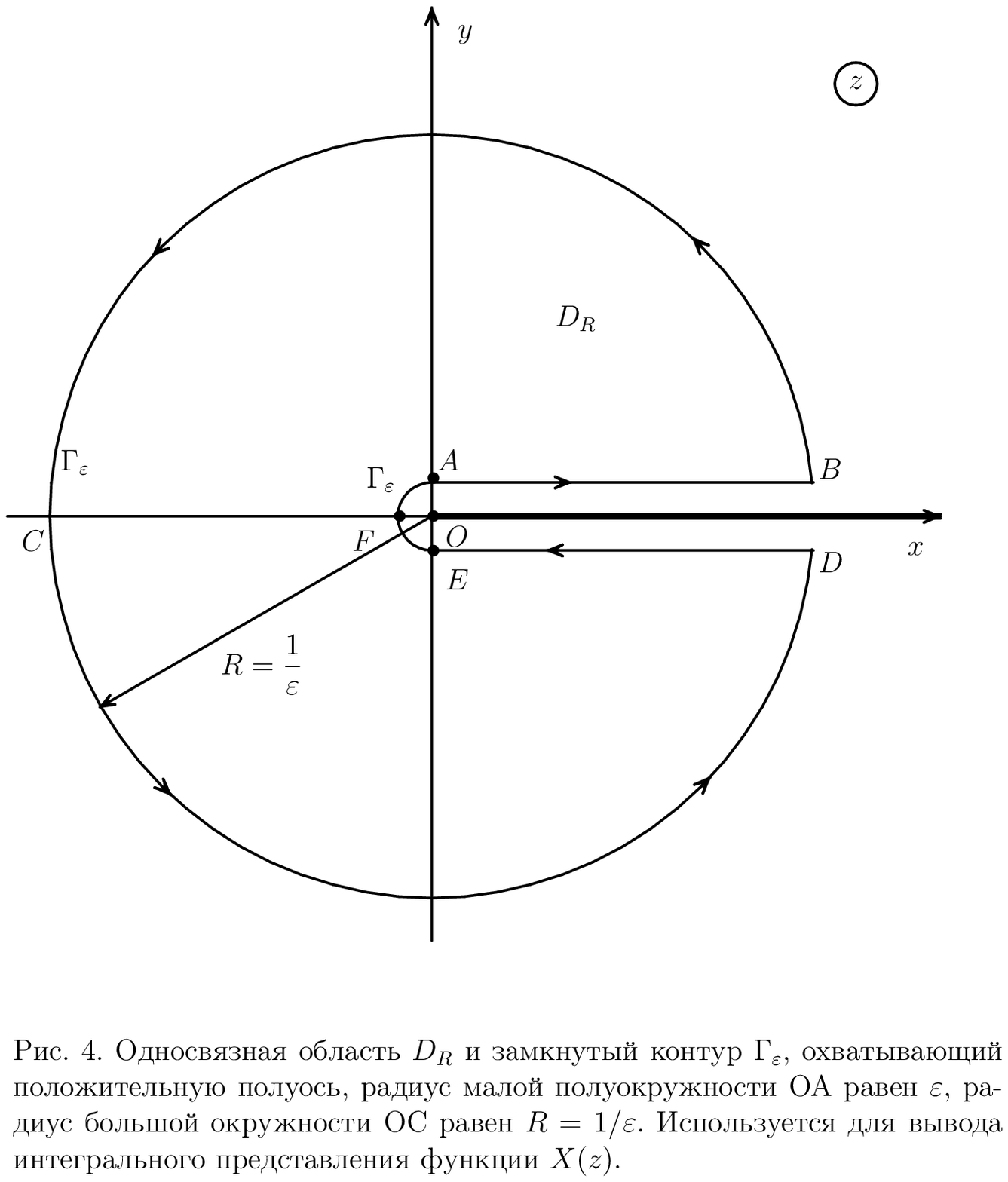}
\end{center}
\end{figure}
\clearpage
$$
\lim_{ \varepsilon \to 0} \int\limits_{BCD}^{}
\dfrac{X(u')}{u'-z}\,du'=0.
\eqno{(4.6)}
$$

Так как функция $X(z)$ ограничена в нуле, то существует $M>0$ такое, что
$|X(u')| \leqslant M$, когда $|u'|=\varepsilon$. Выполним оценку интеграла по полуокружности $EFA$ по модулю. Имеем:
$$
\Bigg|\int\limits_{EFA}^{} \dfrac{ X(u')}{u'-z}\,du'\Bigg|\leqslant
\dfrac{M\pi \varepsilon}{|z|-\varepsilon}.
\eqno{(4.7)}
$$
Теперь мы видим, что из неравенства (4.7) вытекает
равенство (4.5).

Из формул предыдущего п. вытекает, что на
окружности $|z|=R=1/ \varepsilon$ имеем место
следующая оценка $|X(z)|\le M \varepsilon$, где
$M= \exp \{\max\limits_{|z|=R} |\Re V(z)|\}$.
Следовательно, получаем оценку
$$
\left| \int\limits_{BCD}^{} \dfrac{X(z')}{z'-z}\,d z'
\right|\le M \varepsilon \cdot \dfrac{2\pi}{ \varepsilon} \dfrac{1}
{1/ \varepsilon -|z|}= \dfrac{2 \pi M \varepsilon}{1- \varepsilon |z|}.
$$
Из этой оценки и вытекает соотношение (4.6).
Таким образом, переходя к пределу
при $ \varepsilon \to 0$ в равенстве (4.2) и используя
(4.3) и (4.4), получаем следующее интегральное представление
$$
X(z)= \dfrac{1}{2 \pi i} \int\limits_{0}^{\infty}
\dfrac{X^+(u)-X^-(u)}{u-z}\,du.
\eqno{(4.8)}
$$

Это представление выражает значение
функции $X(z)$ в каждой точке $z\in \mathbb{ C \setminus \bar R_+}$
через разность граничных значений этой функции
сверху и снизу на полуоси $ \mathbb{R_+}$.

Приведем ряд родственных к (4.8) интегральных представлений.

Преобразуем интегральное представление (4.8). Для этого воспользуемся краевой задачей (3.1) и представим плотность этого интегрального представления в виде:
$$
\begin{array}{c}
X^+(u)-X^-(u)= X^+(u)- \dfrac{ \lambda^-( u)}{ \lambda^+(
u)}X^+(u)=\\ =X^+(u) \dfrac{ \lambda^+(u)- \lambda^-(u)}{
\lambda^+( u)}=2 \sqrt{ \pi}i\, u e^{-u^2}\dfrac{X^+(u)}{
\lambda^+(u)}.
\end{array}
$$

Тогда интегральное представление (4.8) можно представить в
виде: $$
X(z)= \dfrac{1}{\pi}\int\limits_{0}^{+ \infty} \dfrac{s(u)\,X^+(u)}{
\lambda^+(u)} \dfrac{du}{u-z}.
\eqno{(4.9)}
$$

Интегральное представление (4.9) выражает значение функции $X(z)$ в точке
$z\in \mathbb{C \setminus\bar R_+}$ через ее
граничные значения сверху и снизу на $\mathbb{R_+}$.

Из равенств (3.8) и (3.5) видно, что
$$
\lim\limits_{z \to \infty}zX(z)=1,
$$
а из представления (4.9) следует, что
$$
\lim\limits_{z \to \infty}zX(z)=-\dfrac{1}{\pi}\int\limits_{0}^{+ \infty}
\dfrac{s(u)\,X^+(u)}{ \lambda^+(u)}\,du.
$$

Теперь из последних двух соотношений вытекает, что
$$
\dfrac{1}{\pi}
\int\limits_{0}^{+\infty}\dfrac{s(u)\,X^+(u)}{\lambda^+(u)}\,du= -1.
 $$

Выведем интегральное представление для функции $X(\mu)$
на разрезе $\mu>0$.

На основании интегрального представления (4.8) согласно
формуле Сохоцкого имеем:
$$
X^+(\mu)=\dfrac{X^+(\mu)-X^-(\mu)}{2}+\dfrac{1}{2\pi i}
\int\limits_{0}^{\infty}
\dfrac{X^+(\tau)-X^-(\tau)}{\tau-\mu}\,d\tau,\quad \mu>0,
$$
откуда
$$
\dfrac{X^+(\mu)+X^-(\mu)}{2}=\dfrac{1}{2\pi i}
\int\limits_{0}^{\infty}
\dfrac{X^+(\tau)-X^-(\tau)}{\tau-\mu}\,d\tau,\quad \mu>0,
$$
или,
$$
X(\mu)\ch\Theta(\mu)=\dfrac{1}{\pi i}\int\limits_{0}^{\infty}
\dfrac{X(\tau)\sh\Theta(\tau)d\tau}{\tau-\mu}, \qquad \mu>0.
$$

Выведем интегральное представление функции $X^{-1}(z)$,
необходимое при решении граничных задач. Возьмем функцию
$$
f(z)=\dfrac{1}{X(z)}-z+V_1,
$$
где
$$
V_1=-\dfrac{1}{2\pi i}\int\limits_{0}^{\infty}\ln G(\tau)\,d\tau.
$$

Эта функция имеет асимптотику: $f(z)=O(1/z),\;z\to \infty$.

Проводя те же рассуждения, что и в \cite{19}, получаем:
$$
\dfrac{1}{X(z)}-z+V_1=\dfrac{1}{\pi i} \int\limits_{0}^{\infty}
\dfrac{\sh \Theta(\tau) d\tau}{X(\tau)(\tau-z)},\quad
z\in \mathbb{C\setminus\bar R_+},
\eqno{(4.10)}
$$
и
$$
\dfrac{\ch\Theta(\mu)}{X(\mu)}-\mu+V_1=
-\dfrac{1}{\pi i} \int\limits_{0}^{\infty} \dfrac{\sh\Theta(\tau)d\tau}{X(\tau)(\tau-\mu)},
\quad \mu>0.
\eqno{(4.11)}
$$

Рассмотрим случай нулевого индекса задачи, т.е. $\omega_1\in (\omega_1^*,+\infty)$.

Аналогично предыдущему выводится интегральное представление для точек $z$$\in $$\mathbb{C}\setminus [0,+\infty]$:
$$
X(z)=1+\dfrac{1}{2\pi i}\int\limits_{0}^{\infty}\dfrac{X^+(\mu)-X^-(\mu)}
{\mu-z}d\mu,
$$
или,
$$
X(z)=1+\dfrac{1}{\pi}\int\limits_{0}^{\infty}
\dfrac{s(\tau)X^+(\tau)d\tau}{\lambda^+(\tau)(\tau-z)}.
$$

Из приведенных интегральных представлений нетрудно вывести интегральное
представление на разрезе:
$$
X(\mu)\ch\Theta(\mu)=1+\dfrac{1}{\pi}\int\limits_{0}^{\infty}
\dfrac{s(\tau)X^+(\tau)d\tau}{\lambda^+(\tau)(\tau-\mu)}, \qquad \mu>0.
$$

Для функции $1/X(z)$ точно так же выводим интегральные представления:
$$
\dfrac{1}{X(z)}=1+\dfrac{1}{2\pi i} \int\limits_{0}^{\infty}
\Big[\dfrac{1}{X^+(\tau)}-\dfrac{1}{X^-(\tau)}\Big]\dfrac{d\tau}{\tau-z},
$$
или
$$
\dfrac{1}{X(z)}=1-\dfrac{1}{\pi i}\int\limits_{0}^{\infty}
\dfrac{\sh\Theta(\tau)d\tau}{X(\tau)(\tau-z)},
\qquad z\in \mathbb{C}\setminus \mathbb{R}_+.
\eqno{(4.12)}
$$
Отсюда на разрезе $\mu>0$ получаем следующее интегральное представление:
$$
\dfrac{\ch\Theta(\mu)}{X(\mu)}=1-\dfrac{1}{\pi i}\int\limits_{0}^{\infty}
\dfrac{\sh\Theta(\tau)d\tau}{X(\tau)(\tau-z)}.
\eqno{(4.13)}
$$

\begin{center}
\item{}\section{Факторизация дисперсионной функции}
\end{center}

Здесь устанавливается формула, представляющая
факторизацию дисперсионной функции в верхней и нижней полуплоскостях, а
также выводится формула для факторизации граничных значений
дисперсионной функции сверху и снизу на действительной оси.
Такая факторизация дается в терминах функции $X(z)$.

Пусть сначала $\varkappa(G)=1$, т.е. $\omega_1\in [0,+\infty)$.
Покажем, что для дисперсионной функции $ \lambda(z)$ везде в
комплексной плоскости $ \mathbb{C}$, исключая действительную ось
$ \mathbb{R}$, справедлива формула
$$
\lambda(z)= i\omega_1(z^2-\eta_0^2)X(z)X(-z).
\eqno{(5.1)}
$$

Из этой формулы вытекает, что для ее граничных значений на
$ \mathbb{R}$ справедливы соотношения:
$$
\lambda^{\pm}( \mu)=i\omega_1(\mu^2-\eta_0^2)X^{\pm}( \mu)X(- \mu), \qquad
\mu \ge 0,
\eqno{(5.2)}
$$
$$
\lambda^{\mp}( \mu)=i\omega_1(\mu^2-\eta_0^2)X( \mu)X^{\mp}(- \mu), \qquad
\mu \le 0.
\eqno{(5.3)}
$$

Для доказательства формулы (5.1) введем вспомогательную функцию
$$
R(z)= \dfrac{\lambda(z)}{i\omega_1(z^2-\eta_0^2)X(z)X(-z)}.
\eqno{(5.4)}
$$
Эта функция аналитична везде в комплексной плоскости, кроме точек разрезов $ \mathbb{R_+}$ и $ \mathbb{R_-}$. Точки $z=\pm \eta_0$ являются устранимыми, т.к. в этих точках $\lambda(\pm \eta_0)=0$.

Каждая точка разрезов $ \mathbb{R_+}$ и $ \mathbb{R_-}$ является устранимой. В самом деле, если $ \mu>0$,
то на основании равенства (5.1) и (5.4) имеем:
$$
\dfrac{ \lambda^+( \mu)}{i\omega_1(\mu^2-\eta_0^2)X^+( \mu)X(- \mu)}=
\dfrac{ \lambda^-( \mu)}{i\omega_1(\mu^2-\eta_0^2)X^-( \mu)X( -\mu)},
$$
откуда $ R^+( \mu)= R^-( \mu), \quad \mu>0$.
Если $ \mu<0$, то на основании равенства (5.1), в котором $ \mu$
заменим на $ - \mu$, имеем:
$$
\dfrac{X^+( - \mu)}{X^-( - \mu)}=
\dfrac{ \lambda^+( \mu)}{ \lambda^-( \mu)},
\quad \mu<0.
 $$
Нетрудно видеть, что
 $
\lambda^+( - \mu)= \lambda^-( \mu), \quad \lambda^-( - \mu)=
\lambda^+( \mu).
$
Поэтому
$$
\dfrac{X^+( - \mu)}{X^-(- \mu)}= \dfrac{ \lambda^-( \mu)}
{ \lambda^+( \mu)}, \quad \mu<0,
$$
отсюда
$$
\dfrac{\lambda^+( \mu)}{i\omega_1(\mu^2-\eta_0^2)X( \mu)X^-( - \mu)}=
\dfrac{\lambda^-( \mu)}{i\omega_1(\mu^2-\eta_0^2)X( \mu)X^+( -\mu)}, \quad \mu<0,
$$
или
$$
R^+( \mu)=R^-( \mu ), \quad \mu<0.
$$

Для того чтобы доказать равенства
$$
R^+( \mu)= \dfrac{ \lambda^+( \mu)}{i\omega_1(\mu^2-\eta_0^2)X( \mu)X^-(- \mu)}
$$

\begin{figure}[h]
\begin{center}
\includegraphics[width=15.0cm, height=12cm]{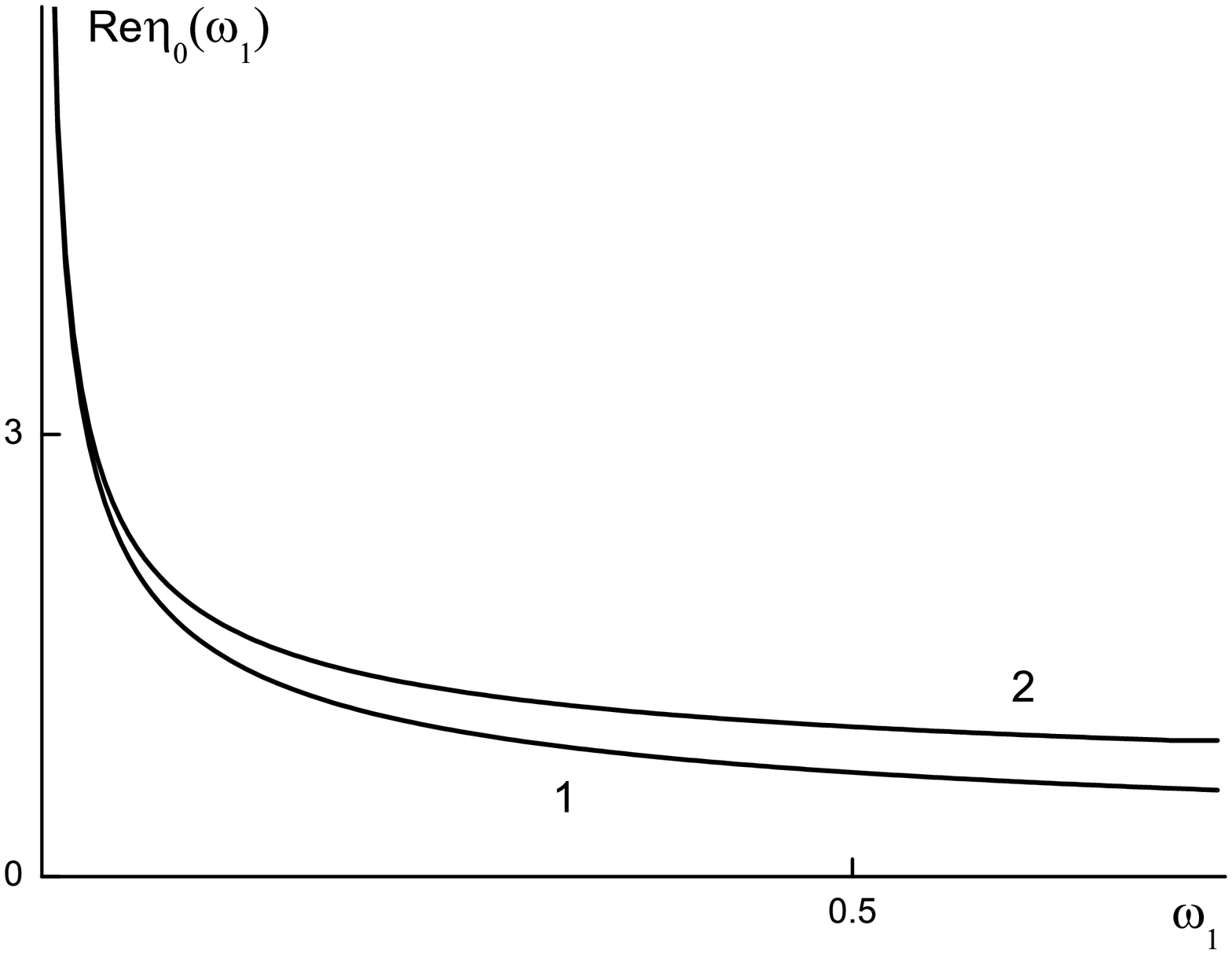}
\end{center}
\begin{center}
{{Рис. 5. Зависимость действительной части нуля дисперсионной функции от параметра $\omega_1$: $\Re \eta_0(\omega_1)$ (кривая $2$), кривая $1$ отвечает действительной части асимптотике этого корня --- $\Re \eta_0^\circ(\omega_1)=\dfrac{1}{2\sqrt{\omega_1}}$.}}
\end{center}
\end{figure}

и
$$
R^-( \mu)=
\dfrac{ \lambda^-( \mu)}{i\omega_1(\mu^2-\eta_0^2)X( \mu)X^+( - \mu)},
$$
заметим, что если точка $z$ стремится к
 точке $ \mu\;( \mu<0)$ из верхней
или нижней полуплоскости, то функции $R^+( \mu)$
или $R^-( \mu)$ вычисляются согласно предыдущим равенствам.
Следовательно, эту функцию можно считать
аналитической функцией везде в $ \mathbb{C}$, в
том числе и в точках разреза, доопределив ее на разрезе по
непрерывности. Осталось заметить, что функция $R(z)$ аналитична
везде в $ \mathbb{ \overline{C}}$ и $R( \infty)=1$.
По теореме Лиувилля эта функция является
тождественно постоянной: $R(z) \equiv 1$,
откуда и вытекает формула (5.1).

Формулы (5.2) и (5.3) очевидно вытекают из формулы (5.1).

Из формулы (5.1) найдем в явном виде формулу для вычисления нулей дисперсионной функции. Для этого вычислим обе части равенства (5.1) в точке $z=i$. В результате для нулей дисперсионной функции получаем следующую формулу:
$$
\eta_0(\omega_1)=\sqrt{-1+\dfrac{i\lambda(i)}{\omega_1}\exp\Big[-V(i)-
V(-i)\Big]}.
$$

\textsc{Замечание 5. 1.} С помощью формулы (5.1) можно упростить интегральное представление (4.9)
$$
X(z)= \dfrac{1}{i\omega_1 \pi} \int\limits_{0}^{\infty}
\dfrac{s(\mu)d\mu}{(\mu^2-\eta_0^2)X(-\mu)(\mu-z)} .
\eqno{(5.5)}
$$

Интегральное представление (5.5) является нелинейным.
Оно выражает значение функции $X(z)$ в точке
$z\in \mathbb{C \setminus R_+}$
через ее значения на отрицательной части действительной оси
$ \mathbb{R_-}$.

Аналог этого представления в теории переноса
излучения используется
для построения функции $X(z)$ численными методами.

Рассмотрим случай нулевого индекса: $\varkappa(G)=0$, т.е. $\omega_1\in [0,\omega_1^*)$. Аналогично предыдущему доказываются формулы
$$
\lambda(z)=-i\omega_1X(z)X(-z), \qquad \Im z \ne 0.
$$
$$
\lambda^{\pm}(\mu)=-i\omega_1X^{\pm}(\mu)X(-\mu), \qquad \mu \leqslant 0.
$$
$$
\lambda^{\pm}(\mu)=-i\omega_1X(\mu)X^{\mp}(-\mu), \qquad \mu <0.
$$

\begin{center}
\bf 6. Заключение 
\end{center}

В настоящей работе развивается математический аппарат для аналитического
решения второй задачи Стокса --- задачи о поведении
разреженного газа, занимающего полупространство над стенкой, совершающей
гармонические колебания. Исследуется краевая задача Римана,
лежащая в основе аналитического решения задачи Стокса. Для решения задачи Римана $X(z)$, называемой факторизующей функцией, и для обратной величины $1/X(z)$ выводится ряд интегральных представлений, необходимых для решения задачи Стокса. Доказываются формулы факторизации дисперсионной функции, также необходимые для решения рассматриваемой задачи Стокса. Выведена формула для вычисления нулей дисперсионной функции.
\newpage
\makeatother {\renewcommand{\baselinestretch}{1.0}

\end{document}